\documentclass[conference]{IEEEtran}
\IEEEoverridecommandlockouts

\usepackage{cite}
\usepackage{csquotes}
\usepackage[many]{tcolorbox}
\tcbuselibrary{theorems}
\usepackage{fontawesome5}
\usepackage{booktabs}
\usepackage{float}
\usepackage{orcidlink}
\usepackage{multirow}
\usepackage{url}
\usepackage{hyperref}
\usepackage{comment}
\usepackage{tabularx}
\usepackage{enumitem}
\usepackage{fancybox}
\usepackage{makecell}
\usepackage{csquotes}
\usepackage{colortbl}

\newtcbtheorem{user}{\faWeixin{} \footnotesize Prompt}
{colback=blue!5,colframe=blue!35!black,fonttitle=\footnotesize \bfseries,fontupper=\footnotesize}{th}

\newtcolorbox{colora}{
enhanced,
boxrule=0pt,frame hidden,
borderline west={2pt}{0pt}{gray!50!black},
colback=gray!05!white,
sharp corners
}

\def\BibTeX{{\rm B\kern-.05em{\sc i\kern-.025em b}\kern-.08em
    T\kern-.1667em\lower.7ex\hbox{E}\kern-.125emX}}
\begin{document}

\makeatletter
\newcommand{\linebreakand}{%
  \end{@IEEEauthorhalign}
  \hfill\mbox{}\par
  \mbox{}\hfill\begin{@IEEEauthorhalign}
}
\makeatother

\title{Automatic Generation of Explainability Requirements and Software Explanations From User Reviews}


\author{
    \IEEEauthorblockN{
        Martin Obaidi\orcidlink{0000-0001-9217-3934}, 
        Jakob Droste\orcidlink{0000-0001-8746-6329}, 
        Hannah Deters\orcidlink{0000-0001-9077-7486}, \\
        Marc Herrmann\orcidlink{0000-0002-3951-3300}, 
        Kurt Schneider\orcidlink{0000-0002-7456-8323}
    }
    \IEEEauthorblockA{
        \textit{Leibniz Universität Hannover, Software Engineering Group} \\
        Hannover, Germany \\
        \{martin.obaidi, jakob.droste, hannah.deters\}@inf.uni-hannover.de, \\
        \{marc.herrmann, kurt.schneider\}@inf.uni-hannover.de
    }
    \and
    \IEEEauthorblockN{
        Jil Klünder\orcidlink{0000-0001-7674-2930}
    }
    \IEEEauthorblockA{
        \textit{University of Applied Sciences, FHDW Hannover} \\
        Hannover, Germany \\
        jil.kluender@fhdw.de
    }
    \and
    \IEEEauthorblockN{
        Hugo Villamizar
    }
    \IEEEauthorblockA{
        \textit{fortiss GmbH} \\
        Munich, Germany \\
        guarinvillamizar@fortiss.org
    }
    \and
    \IEEEauthorblockN{
        Jannik Fischbach\orcidlink{0000-0002-4361-6118}
    }
    \IEEEauthorblockA{
        \textit{Netlight Consulting GmbH and fortiss GmbH} \\
        Munich, Germany \\
        jannik.fischbach@netlight.com
    }
    \and
    \IEEEauthorblockN{
        Steffen Krätzig
    }
    \IEEEauthorblockA{
        \textit{Phoenix Contact GmbH \& Co. KG} \\
        Blomberg, Germany \\
        steffen.kraetzig@phoenixcontact.com
    }
}

\maketitle

\begin{abstract}
Explainability has become a crucial non-functional requirement to enhance transparency, build user trust, and ensure regulatory compliance. However, translating explanation needs expressed in user feedback into structured requirements and corresponding explanations remains challenging. While existing methods can identify explanation-related concerns in user reviews, there is no established approach for systematically deriving requirements and generating aligned explanations. To contribute toward addressing this gap, we introduce a tool-supported approach that automates this process. To evaluate its effectiveness, we collaborated with an industrial automation manufacturer to create a dataset of 58 user reviews, each annotated with manually crafted explainability requirements and explanations. Our evaluation shows that while AI-generated requirements often lack relevance and correctness compared to human-created ones, the AI-generated explanations are frequently preferred for their clarity and style. Nonetheless, correctness remains an issue, highlighting the importance of human validation. This work contributes to the advancement of explainability requirements in software systems by (1) introducing an automated approach to derive requirements from user reviews and generate corresponding explanations, (2) providing empirical insights into the strengths and limitations of automatically generated artifacts, and (3) releasing a curated dataset to support future research on the automatic generation of explainability requirements.
\end{abstract}


\section{Introduction}
\textbf{Context.} Modern software systems are becoming increasingly complex and difficult for users to understand, largely due to the incorporation of artificial intelligence (AI) and data-driven components. This has led to a growing demand for greater transparency and comprehensibility~\cite{adadi2018peeking}. As a result, \textit{explainability} has emerged as a critical non-functional requirement for software systems to foster user trust and meet regulatory requirements~\cite{kohl_explainability_2019,brunotte_quo_2022}. Explainability is often considered a means to achieve transparency, as explanations provide insights into system behavior~\cite{deters2024qualitymodel,Deters2025quality,deters2025identifying}. However, it also serves broader purposes, such as increasing trust, efficiency, or usability~\cite{deters2024qualitymodel,Deters2025quality}. Conversely, transparency can also be enhanced through mechanisms other than explainability, such as open system documentation or interpretable user interfaces.

Despite its importance, developing \textit{explainable} systems remains challenging for several reasons. First, users’ needs for explanation are often expressed in informal, subjective ways, such as complaints or vague remarks, which makes them difficult to identify and formalize. Second, translating these explanation needs into structured requirements is non-trivial, as it requires reconciling subjective user perceptions with established quality criteria for requirements~\cite{Frattini22}. Third, even when suitable explanations are created, they must be integrated into the system in ways that align with user context and expectations; poorly placed or irrelevant explanations can harm the user experience~\cite{Chazette2020}.

\textbf{Research Gap.} Prior work has addressed explainability from a requirements engineering (RE) perspective by focusing on the identification and classification of explanation needs. For example, user reviews have been explored as a valuable source for surfacing implicit demands for clarity and transparency~\cite{Unterbusch23,anders2023userfeedback,anders2022userfeedback}, and taxonomies have been proposed to categorize these needs across different contexts and user expectations~\cite{droste2024explanations,Unterbusch23}. While these contributions provide valuable insights into what users may find unclear, they do not address how such needs can be systematically translated into structured requirements or implemented through suitable explanations. This landscape highlights that, unlike more established non-functional requirements such as performance or security, explainability requirements require more systematic attention regarding their operationalization.

\textbf{Principal Idea.} In this work, we propose a tool-supported approach that transforms explanation needs expressed in user reviews into structured explainability requirements and corresponding explanations. Building on the idea that user reviews provide valuable insights into user reasoning and perceptions of system behavior~\cite{henao21}, our approach automates both the derivation of requirements and the generation of explanations. For evaluation, we applied it to user reviews of the Spotify mobile app, which offers a rich set of real-world explanation needs across diverse user interactions. To support validation, four experienced requirements engineers from an industrial automation company reviewed the generated requirements and explanations based on quality criteria including clarity, correctness, relevance, and level of detail. 

\textbf{Contributions.} Our study results show a general preference for manually created explainability requirements, particularly in terms of relevance and correctness. In contrast, preferences for explanations were mixed: participants appreciated the clarity and tone of some AI-generated explanations, but concerns around correctness and alignment with user needs persisted. These findings suggest that while automation shows promise, human validation remains essential, especially for deriving high-quality explainability requirements. The main contributions of this work are as follows:
\begin{itemize}
    \item An automated approach for deriving explainability requirements from user reviews and generating corresponding explanations, supporting the operationalization of explainability as a non-functional requirement in software systems. 
    \item A curated dataset of 58 user reviews annotated with both manually derived and automatically generated explainability requirements and explanations, serving as a benchmark for future research.
    \item Empirical insights into the strengths and limitations of automatically generated explainability artifacts, based on a comparative study involving requirements engineers and end users.
\end{itemize}

The paper is structured as follows: Section~\ref{sec:background} presents related work and the theoretical background. Section~\ref{sec:automation} describes our approach for the automated generation of explainability requirements and explanations. Section~\ref{sec:study} details the study design, including data collection and methodology. The results of our evaluation are presented in Section~\ref{sec:results}. In Section~\ref{sec:discussion}, we analyze and interpret the findings, discuss limitations, address threats to validity, and outline directions for future research. Finally, Section~\ref{sec:conclusion} concludes the paper.

\textbf{Data Availability.} All study data and our code are publicly available at \href{https://doi.org/10.5281/zenodo.15839752}{Zenodo}~\cite{obaidi2025datasetexplainability}.

\section{Background and Related Work}
\label{sec:background}

\textbf{Explanations in Software.}
Explaining systems to end-users has long been a topic of interest in research on human-computer interaction~\cite{kieras1984role,staggers1993mental,obaidi2025elicit}. Past work has identified explanations in software as a suitable means to enable accurate mental models in users~\cite{sokol2020one,tankelevitch2024metacognitive}. As such, explanations may guide users through interactions with the software, increasing ease of use~\cite{droste2024explanations,obaidi2025mood,obaidi2025appKonwledge,obaidi2025AppFeaturesExplainNeeds}.
With the rise of AI over the last decades, a significant body of research into software explanations has focused on eXplainable Artificial Intelligence~(XAI)~\cite{adadi2018peeking}. In XAI, explanations are mainly used to provide transparency and understandability concerning system behavior~\cite{droste2024peeking}, which also contributes to trustworthiness~\cite{kastner2021relation}. 

We follow a holistic understanding of explanations in software, meaning that explanations can go beyond reasoning about system behavior, which is common in XAI~\cite{droste2024peeking}. Instead, we acknowledge that explanations may also address other aspects of a system, such as navigation, operation, or privacy concerns. Recent work in RE supports this perspective~\cite{droste2024explanations,brunotte_quo_2022}. 
Providing inappropriate explanations, too many explanations, or missing the right timing can hinder user experience rather than enhance it~\cite{chazette2020explainability,deters2024x}. If implemented incorrectly, explanations may lead to unnecessary cognitive load~\cite{nunes2017systematic}, resulting in confusion and frustration among end-users. Therefore, it is crucial to elicit explainability requirements carefully to ensure that explanations are provided appropriately and at the right time.


\textbf{Explainability Needs in End-User Feedback.}
Droste \textit{et al}.~\cite{droste2024explanations} conducted an online survey to collect feedback from 84 software users to establish a taxonomy for explainability needs in everyday software systems. Their findings indicate that the need for explaining system behavior, which is commonly researched in XAI, also exists in everyday software systems. However, this need ranked second to the need for explanations regarding interactions between the user and the system. This aligns more closely with the work of Kieras and Bovair~\cite{kieras1984role}, who discovered that explaining interactions with a system enables users to operate it more efficiently. In addition to interaction and behavior explanations, Droste \textit{et al}.~\cite{droste2024explanations} identified needs for explanations on domain knowledge, privacy and security, and user interfaces. Unterbusch \textit{et al}.~\cite{Unterbusch23} collected user feedback from 1,730 app reviews to identify and categorize explainability needs. Their approach differentiates between user feedback explicitly requesting explanations and feedback that describes problems that could be mitigated by explanations. Obaidi \textit{et al}.~\cite{obaidi2025automatingexplanationneedmanagement} analyzed explanation needs in app reviews from a company specializing in navigation apps. They proposed an automated method to detect explanation needs and identify the most suitable team or source within the company to provide the required information.

While prior work has addressed the identification and categorization of explanation needs in user feedback~\cite{droste2024explanations,Unterbusch23}, as well as methods for extracting such needs automatically, there is limited research on systematically deriving structured explainability requirements from these needs and generating actionable explanations for end users. Our work addresses this research gap.

\textbf{Large Language Models in RE.}
Vogelsang and Fischbach~\cite{vogelsang2024using} discuss the automation of RE tasks using Large Language Models (LLMs). On a high level, the authors differentiate between tasks that involve understanding and those that involve generation. Understanding tasks focus on analyzing common RE artifacts, such as classifying requirements into types~\cite{hey2020norbert} or applying sentiment analysis to user feedback~\cite{guzman2014users,obaidi2025TrustworthySA,obaidi2025GoldStandardDE,obaidiSentiSMS22,obaidi2021development,obaidi22cross}. In contrast, generation tasks focus on creating new artifacts, such as generating test cases from requirements~\cite{fischbach2023automatic} and synthesizing new requirements~\cite{koscinski2023demand}.

The recent trend of using large language models in RE also includes research on prompt engineering and optimization~\cite{vogelsang2024using}, but our study employs practical, template-based prompts as typically used in RE practice, focusing on feasibility and replicability.

\section{Automatic Creation of Explainability Requirements and Explanations} \label{sec:automation}
To automate the creation of explainability requirements and corresponding explanations, we developed a prompt-based approach that interprets user reviews expressing explanation needs. The underlying model is expected to handle multiple tasks: natural language understanding, requirements formulation, and fluent text generation. Although sentiment analysis is not explicitly required, it can help contextualize emotionally charged or ambiguous feedback. Given their strong performance in context-aware language tasks, Generative Pretrained Transformer (GPT) models are well-suited for this setting. We used ChatGPT 3.5—the latest available model at the time of the study—to generate one explainability requirement and one explanation for each selected user review.

\subsection{Deriving Explainability Requirements from Explanation Needs}
As a first step, we adapted the model to interpret explanation needs and derive requirements. In initial experiments, we provided the model with the full user review and prompted it to generate an explainability requirement. 
Our goal was to establish the prompt in an ad-hoc, practical manner rather than through systematic prompt engineering, as this approach is common in practice. Practitioners often iterate and adapt prompts to their immediate needs rather than conducting extensive prompt studies in advance.

We found that although the model followed the desired structure, it sometimes included superfluous or emotional content, as user reviews often contain informal or subjective language. To address this, we instructed ChatGPT to first summarize the review—removing unnecessary or emotional content—before further processing. This pre-processing step allowed us to distill the essential information for the final requirement and ensured concise output. One author manually reviewed all 58 summaries to confirm their accuracy and completeness, ensuring each captured the relevant aspects of the explanation need while filtering out extraneous content.

We used the following prompt for the pre-processing step, with the temperature set to 0.5 to balance creativity and adherence to the rules.

\begin{user*}{Pre-Process and Summarize Review}
Do not use headings; use only plain text. Summarize the following review, focusing on the essentials. Remove superfluous and emotional content. Keep ambiguity on the user's side. Use German.
\end{user*}

To derive the explainability requirements, we used a low temperature value of 0.3 to encourage the model to generate formal, consistent outputs. The results produced by ChatGPT with this prompt were highly consistent, with only minor variations.

\begin{user*}{Derive Explainability Requirement}
You are an expert at writing explainability requirements. Create a short explainability requirement (if several explainability requirements are necessary, then several) for the system developers from the following review. Pay attention to the desired properties of a good requirement. Remove any emotions or opinions of the user. Avoid the first-person perspective. Use the sentence structure \enquote{The system must explain...} when formulating the requirement. Only state the requirement and keep the requirement objective.
\end{user*}

\subsection{Generating Explanations}
Formulating effective explanations may require domain-specific knowledge about the application's functionality. We hypothesize that ChatGPT,
through its extensive pre-training, has enough domain knowledge to suggest a
possible explanation text. While the model is capable of producing coherent outputs even with limited context, we acknowledge that certain explanations may require refinement when factual accuracy is critical. For this task, we used a temperature setting of 0.8 to encourage more natural and creative phrasing, aiming to align the generated text with typical user-facing language styles.

\begin{user*}{Generate Explanation Text}
Write a short explanation that should be shown in the UI to fulfill the following explainability requirement.
\end{user*}

In line with common software development practices, requirements are defined before implementing corresponding functionalities. We therefore instructed ChatGPT to generate explanations based on the previously derived requirements, not directly from review summaries. While summaries condense user feedback and remain relatively fixed, explainability requirements enable targeted refinement and adjustment. By modifying the focus of the requirement, we ensure that the generated explanation is formulated appropriately and addresses the user's need more directly.

The final output of our approach consists of the derived explainability requirement and a suitable explanation to be incorporated into the UI.

\subsection{Tool Support}
To support development teams in applying our approach, we built a tool that implements all described steps. The tool consists of a web-based frontend and a backend that sends prompts and user input to the ChatGPT API (frontend: Vue.js; backend: Python/Flask). As shown in Figure~\ref{fig:tool-example}, users can either manually enter reviews or upload a CSV file. Each step—summarization, requirement derivation, explanation generation—can be performed individually, and all outputs can be exported for further use.

\begin{figure*}[ht]
    \centering
    \includegraphics[width=1\linewidth]{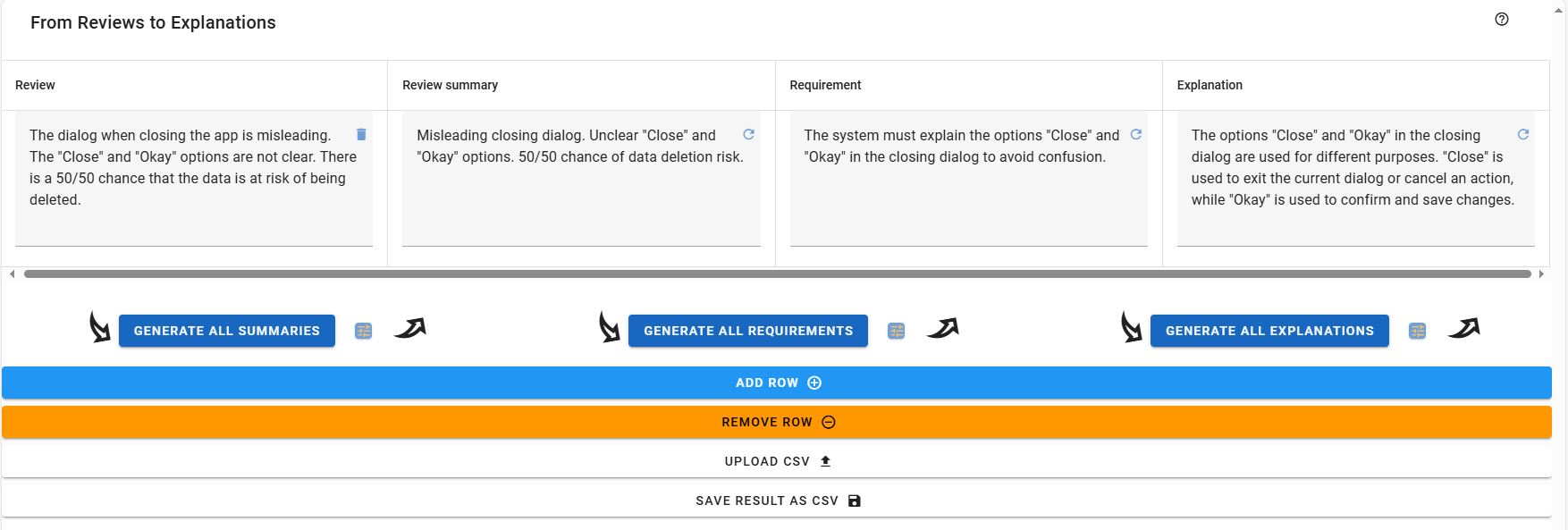}
    \caption{Overview of our tool-supported approach for automated creation of explainability requirements and explanations.}
    \label{fig:tool-example}
\end{figure*}

\section{Study Design}
\label{sec:study}
Our study aims to evaluate the quality of automatically generated explainability requirements and explanations in comparison to those manually crafted by requirements engineers. The study focuses on explanation needs extracted from user reviews of the Spotify mobile app~\cite{obaidi-goldstandard-explain-zenodo2024}. Based on these reviews, two sets of artifacts were produced for each explanation need: one set manually created by requirements engineers in a dedicated workshop, and another automatically generated using our tool-supported approach. Each set consisted of an explainability requirement and a corresponding explanation.

To compare the quality of these artifacts, we conducted two separate online surveys. In the first, requirements engineers were asked to assess and select their preferred explainability requirement for each review. In the second, end users evaluated the explanations and selected the version they found clearer or more helpful. To ensure consistency, each explanation need was associated with exactly one manually created and one AI-generated requirement and explanation, resulting in a one-to-one mapping across the dataset.

\subsection{Research Goal and Research Questions}

We defined our study goal using the Goal–Question–Metric (GQM) template~\cite{basili2002tame} as follows:\\

\setlength{\shadowsize}{2pt}
\noindent
\shadowbox{
\begin{minipage}[t]{0.95\columnwidth}
\textbf{Research Goal:} \textit{Analyze} the effectiveness of automatically generated explainability requirements and explanations  
\textit{for the purpose of} understanding how they compare to manually crafted ones in terms of perceived preference and quality  
\textit{with respect to} clarity, correctness, relevance, level of detail, style, and tone  
\textit{from the point of view of} requirements engineers and end users  
\textit{in the context of} a controlled study involving expert review and user feedback.
\end{minipage}
}\vspace{0pt}

Based on our goal, we formulated two research questions:

\begin{itemize}
\item \textbf{RQ1: How do requirements engineers perceive and prefer AI-generated explainability requirements compared to manually created ones?} 
This question investigates whether AI-generated requirements meet the quality standards and expectations of experienced requirements engineers, and identifies potential shortcomings in automatically generated requirements.

\item \textbf{RQ2: How do users perceive and prefer AI-generated explanations compared to manually created ones?} 
This question explores whether AI-generated explanations fulfill user expectations regarding clarity, correctness, and relevance, key factors for fostering transparency and trust.
\end{itemize}


\subsection{Workshop: Manual Formulation of Explainability Requirements and Explanations }\label{sec:corpus}
To the best of our knowledge, this work is the first to report on an approach capable of automatically deriving explainability requirements from explanation needs and proposing suitable explanations. Consequently, no established dataset was available that could be used to evaluate such an approach.

\textbf{Data Collection.} To answer RQ1 and RQ2, we require a dataset of user reviews that includes explanation needs, a corresponding explainability requirement for each review, and a suitable explanation that implements the requirement. We draw on a dataset~\cite{obaidi-goldstandard-explain-zenodo2024}, which analyzed 4,495 out of 90,000 user reviews regarding the occurrence of explanation needs. In total, the dataset includes 2,186 reviews, each containing a need for an explanation. 

To create our dataset, we collaborated with our industry partner, a manufacturer of industrial automation, and focused on user reviews that describe explanation needs related to using the Spotify app, which includes 139 reviews marked accordingly. Specifically, we shuffled these reviews and randomly selected 58 reviews, ensuring that all types and categories of explanation needs (such as \textit{explicit}, \textit{implicit}, and various thematic categories) were represented according to the taxonomy by Droste \textit{et al}.~\cite{droste2024explanations}. 

The Spotify app was chosen because it is a widely used application with a large pool of relevant user reviews. Additionally, the requirements engineers from our industry partner regularly use the app and are familiar with its functionality and typical user concerns. This familiarity enabled them to accurately interpret explanation needs and derive suitable requirements and explanations, while still remaining impartial, as they were not involved in the app's development.

\textbf{Participants' Demographics.} The workshop included four requirements engineers from the industrial automation manufacturer. Their levels of experience with RE varied: one participant reported working with requirements weekly, two engaged with them daily, and one interacted with requirements multiple times per day.

\textbf{Workshop Procedure.}
Participants were first introduced to the concept of explainability through examples of explanation needs, explainability requirements, and explanations. Given their extensive experience in RE, no basic introduction to the field was necessary. During the workshop, the requirements engineers collaboratively derived explainability requirements based on the explanation needs expressed in the 58 selected Spotify reviews. To facilitate this process, we provided a structured Excel sheet where participants reviewed each user review and documented the corresponding requirement. After formulating the requirements, the engineers created explanations that addressed the identified needs. Requirements and explanations were recorded in the sheet during the session.



\subsection{Online Surveys: Voting on Explainability Requirements and Explanations}
\label{sec:abstimmen}

\textbf{Study Objects and Participants.}  
We used the dataset described in Section~\ref{sec:corpus}, which comprises 58 Spotify reviews, each with both manually crafted and AI-generated explainability requirements and corresponding explanations. For RQ1, eight requirements engineers from our industry partner participated in the evaluation. Importantly, none of them were involved in the dataset creation, which helps minimize bias and ensures an independent assessment. For RQ2, we recruited 14 end users via social networks, friends, and family. Since explanations are ultimately intended for general users rather than RE professionals, these participants were well suited to evaluate which version of each explanation they found clearer or more appropriate, without requiring detailed knowledge of Spotify’s internal functionality.

\textbf{Study Approach.}  
The evaluation followed a comparative design: For each review, participants were presented with both the manually and automatically generated artifacts (in random order and blinded to their origin) and selected their preferred requirement (RQ1) or explanation (RQ2). In both studies, participants were unaware of the origin of each artifact to ensure an unbiased assessment. To measure interrater agreement, we calculated Fleiss' Kappa for both experiments.

Participants were also asked to justify their selections. They could choose from a predefined set of reasons (allowing multiple selections) and/or provide free-text justifications. These open responses were later categorized by the first two authors and mapped to existing or emergent reason categories, as appropriate. The evaluation focused on six quality criteria:

\begin{enumerate}
    \item Tone: Appropriateness of tone (\textit{e.g.}, friendly, supportive, neutral)
    \item Style: Appropriateness of linguistic style (\textit{e.g.}, formal, colloquial)
    \item Clarity: Ease of understanding, simple formulation, absence of technical jargon or ambiguity
    \item Correctness: Factual accuracy, or perceived correctness if domain knowledge is lacking
    \item Relevance: Directness in addressing the user's concern, avoidance of irrelevant information
    \item Level of detail: Appropriateness of the amount of information provided
\end{enumerate}

These criteria were chosen to provide a structured and comprehensive assessment of linguistic and content quality, based on established standards in RE and explainability research.

To further assess the tool's practical applicability, we conducted an additional study with the same eight requirements engineers, comparing the time required for manual requirement formulation to that needed when using the tool. They also rated their confidence in the generated requirements and the perceived usefulness of the tool. These results provide insights into the efficiency and real-world value of the tool in RE workflows.

\section{Results} \label{sec:evaluation}
\label{sec:results}

\subsection{Answering RQ1: Explainability Requirement Evaluation}
\label{sec:RequirementEvaluation}

For the evaluation of generated explainability requirements, we involved eight requirements engineers from our industry partner. The participants’ ages ranged from 28 to 44 years (median 34.5, average 35.1); the group consisted of seven male and one female engineer. Participants had between 2 and 19 years of experience in RE (median 8, average 8.5 years). Regarding their frequency of working with requirements: two did so weekly, four daily, and two several times a day.

The results indicate a clear, but not absolute, preference for manually created requirements. Across all reviews, ChatGPT-generated requirements were chosen 168 times, whereas manually created requirements were selected 296 times. For 28 of the 58 reviews, all participants unanimously agreed: in 21 cases, the manual requirement was preferred, in 7 cases the ChatGPT version. This corresponds to an overall agreement rate of 48\% and a Fleiss’ Kappa value of $\kappa = 0.49$, which indicates moderate agreement.

In the majority of reviews (32 out of 58), most participants preferred the manually created requirement. For 14 reviews, the AI-generated requirement was favored by the majority. In 12 reviews, the votes were split, with no clear majority for either option (see Figure~\ref{fig:evaluation-requirement}).

\begin{figure}[ht]
    \centering
    \includegraphics[width=1\linewidth]{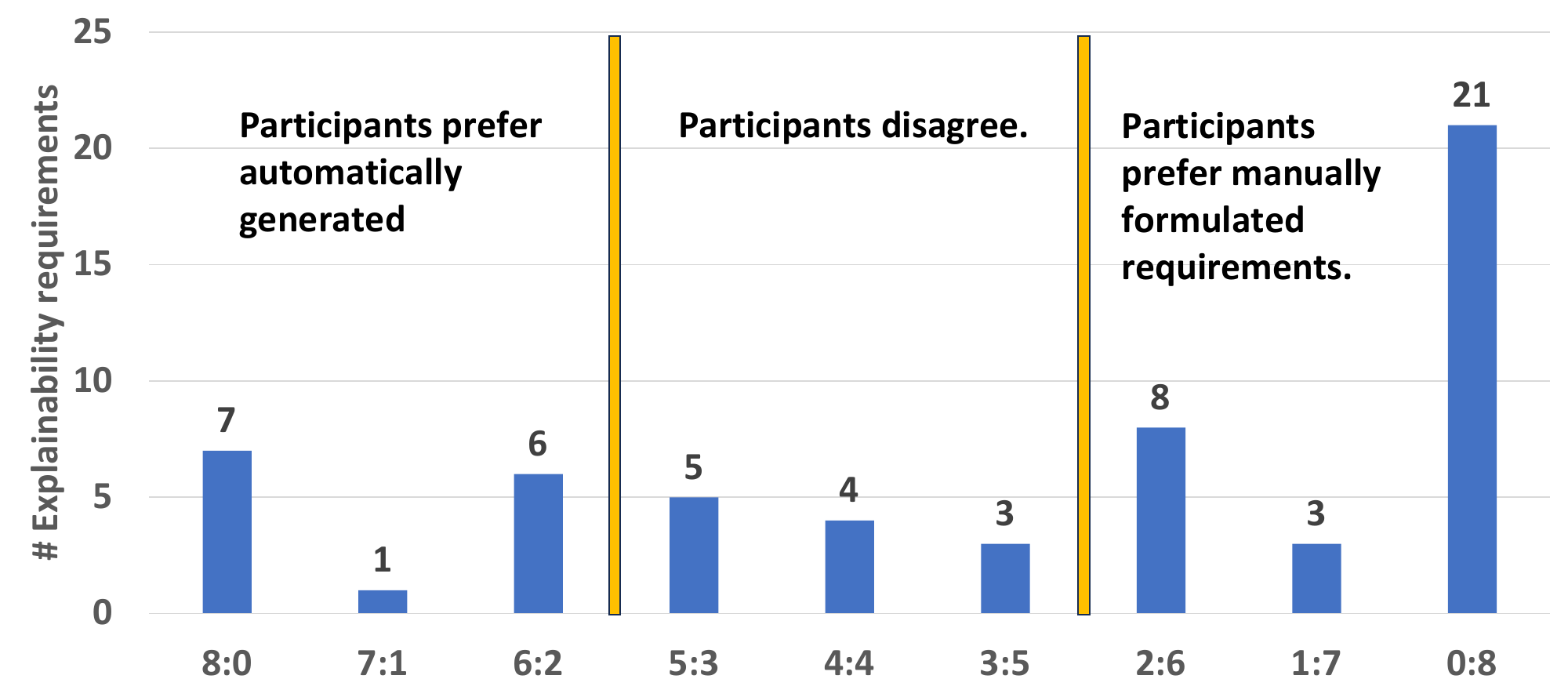}
    \caption{Distribution of decisions by eight requirements engineers for preferred explainability requirements: ChatGPT-generated vs. manually created (ratio: ChatGPT:manual).}
    \label{fig:evaluation-requirement}
\end{figure}

To further analyze participants' decision-making, we examined the evaluation criteria cited as reasons for their choices. As Table~\ref{tab:evaluation_criteria} shows, the most frequent reasons were Relevance (252 mentions) and Correctness (194 mentions), followed by Clarity (71) and Level of Detail (84). Style and Tone were rarely considered decisive. In cases of strong disagreement (ties), the most common combinations of criteria were Correctness and Relevance (six times), followed by Clarity and Relevance (three times), Level of Detail and Correctness (once), and Correctness and Clarity (once).

\begin{table}[ht]
    \centering
    \caption{Summary of Evaluation Criteria Counts for Explainability Requirements}
    \begin{tabularx}{\columnwidth}{l >{\raggedleft\arraybackslash}X >{\raggedleft\arraybackslash}X >{\raggedleft\arraybackslash}X >{\raggedleft\arraybackslash}X >{\raggedleft\arraybackslash}X >{\raggedleft\arraybackslash}X}
        \toprule
         & \textbf{Detail level} & \textbf{Correct.} & \textbf{Clarity} & \textbf{Style} & \textbf{Relevance} & \textbf{Tone} \\
        \midrule
        \textbf{Occurrence} & 18 & 39 & 18 & 0 & \textbf{51} & 1 \\
        \textbf{Sum} & 84 & 194 & 71 & 0 & \textbf{252} & 6 \\
        \textbf{Majority} & 7 & \textbf{18} & 3 & 0 & \textbf{18} & 1 \\
        \bottomrule
    \end{tabularx}
    \label{tab:evaluation_criteria}
\end{table}

To assess whether the observed preference distribution deviates from random choice, a Chi-Square test was performed. Assuming a 50:50 split as baseline, the result (Chi$^2 = 35.31$, p = 2.81e-09) indicates a statistically significant preference for manually created requirements. This is well below the standard significance thresholds ($p < 0.05$ and $p < 0.01$), providing strong evidence that participants did not choose at random but clearly favored manual requirements.

\begin{table}[ht]
    \centering
    \caption{Chi-Square Test Results for Explainability Requirement Preferences}
    \begin{tabularx}{\columnwidth}{l r r r r}
        \toprule
        \textbf{Category} & \textbf{\# ChatGPT} & \textbf{\# Manual} & \textbf{Chi$^2$} & \textbf{p-value} \\
        \midrule
        Relevance & 80  & 98  & 0.03  & 0.863 \\
        Correctness & 64  & 87  & 0.02  & 0.890 \\
        Clarity & 62  & 86  & 0.00  & 0.986 \\
        Level of Detail & 36  & 70  & 1.48  & 0.223 \\
        Tone & 14  & 26  & 0.04  & 0.840 \\
        Style & 10  & 21  & 0.05  & 0.822 \\
        \bottomrule
    \end{tabularx}
    \label{tab:chi_square_results}
\end{table}

To investigate whether the reasons for preferring ChatGPT-generated versus manually created requirements differ systematically, we applied Chi-Square tests to the distribution of reasons (Table~\ref{tab:chi_square_results}). None of the differences were statistically significant (all $p > 0.05$). The most frequently cited reason in both groups was Relevance, followed by Correctness and Clarity. Overall, the results show a statistically significant preference for manually created requirements, primarily justified by perceived relevance and correctness. However, there is no statistically significant difference in the types of criteria cited for the two artifact sources. The moderate Fleiss’ Kappa suggests some variability in individual preferences, highlighting the subjective nature of evaluating requirements quality.

\subsection{Answering RQ2: Explanation Evaluation}

The explanation evaluation involved 14 participants (12 male, 2 female), aged 22–39 years (median: 26.5, average: 27.4). Five were students, one was employed, seven combined work and study (\textit{e.g.}., doctoral candidates), and two specified their roles in research and teaching. Music app usage among participants was high: six used music apps daily, another six several times per week, one several times a month, and one a few times per year; none reported never using such apps.

In total, participants chose ChatGPT-generated explanations 365 times and manually created explanations 447 times. However, strong disagreement was typical: for 31 of the 58 reviews, no clear majority emerged (see Figure~\ref{fig:overview-approach}). Only in three cases did all participants agree, always in favor of the manually created explanation. For 9 reviews, the AI-generated explanation was preferred by most, and for 18 reviews, the manual explanation was favored. The overall agreement rate was low (5\%), with a Fleiss’ Kappa of $\kappa = 0.16$, indicating only slight agreement.

\begin{figure}[ht]
    \centering
    \includegraphics[width=1\linewidth]{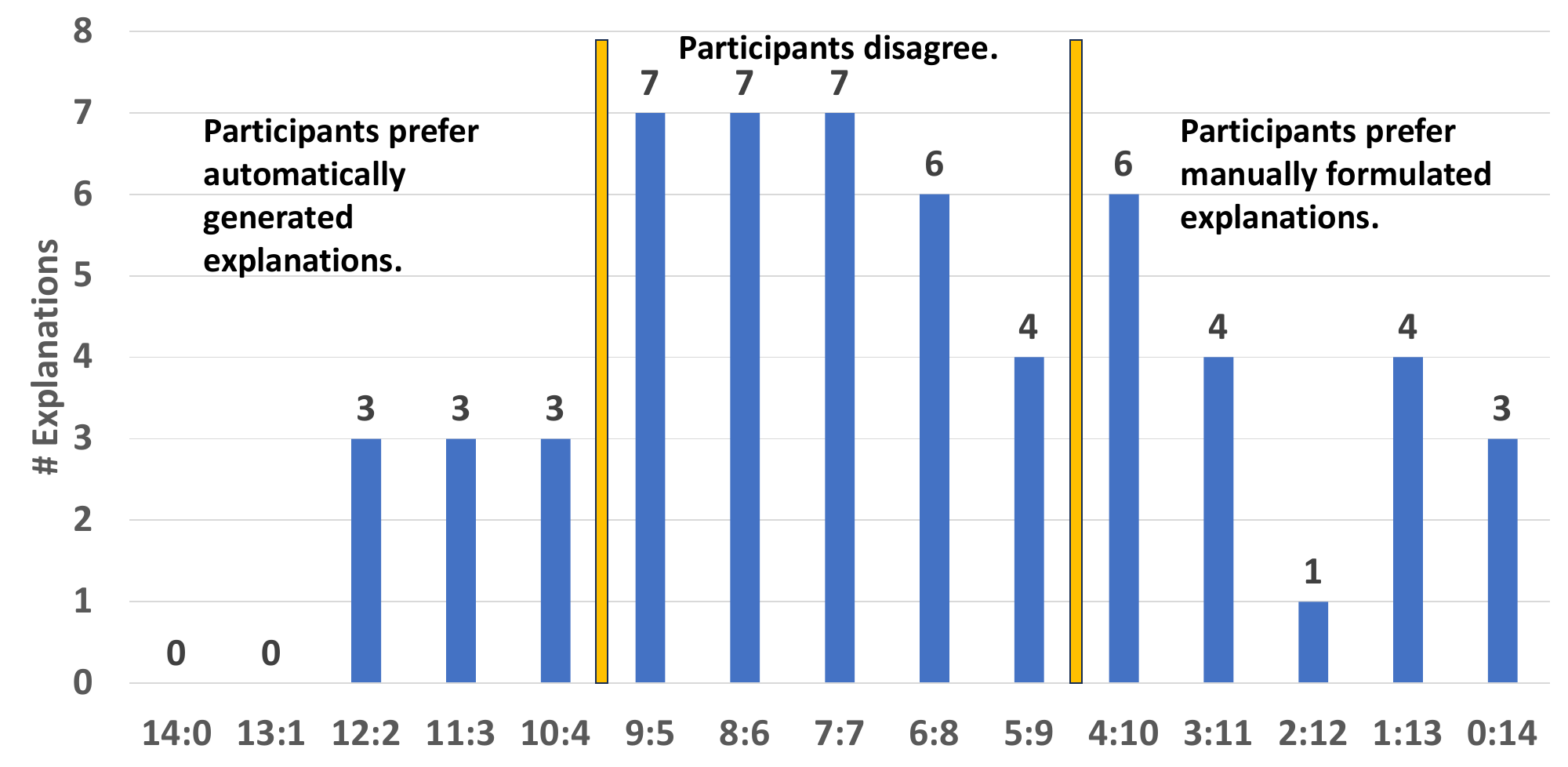}
    \caption{Distribution of decision ratios by 14 users on their preferred explanation, comparing ChatGPT-generated versus manually created explanations (ChatGPT:manual).}
    \label{fig:overview-approach}
\end{figure}

Analysis of evaluation criteria (Table~\ref{tab:evaluation_explanations}) shows that Correctness and Relevance were most frequently cited as decisive, each mentioned 58 times as the main reason for a decision. Clarity, Level of Detail, and Style were also frequently mentioned, while Tone was the least relevant. In reviews where no majority was reached, the most common tie was between Correctness and Relevance.

\begin{table}[ht]
    \centering
    \caption{Summary of Evaluation Criteria Counts for Explanations}
    \begin{tabularx}{\columnwidth}{l >{\raggedleft\arraybackslash}X >{\raggedleft\arraybackslash}X >{\raggedleft\arraybackslash}X >{\raggedleft\arraybackslash}X >{\raggedleft\arraybackslash}X >{\raggedleft\arraybackslash}X}
        \toprule
         & \textbf{Detail level} & \textbf{Correct.} & \textbf{Clarity} & \textbf{Style} & \textbf{Relevance} & \textbf{Tone} \\
        \midrule
        \textbf{Occurrence} & 48 & \textbf{58} & 56 & 53 & \textbf{58} & 26 \\
        \textbf{Sum} & 88 & \textbf{358} & 164 & 144 & 178 & 44 \\
        \textbf{Majority} & 1 & \textbf{41} & 3 & 6 & 1 & 1 \\
        \bottomrule
    \end{tabularx}
    \label{tab:evaluation_explanations}
\end{table}

A Chi-Square test was performed to determine whether preferences for manually and automatically generated explanations deviated from random choice (50:50 baseline). The result (Chi$^2$ = 8.28, $p = 0.004$) indicates a statistically significant deviation from an even split ($p < 0.05$), although no clear trend toward either manual or AI explanations emerged in the overall result. A further Chi-Square analysis tested whether the distribution of evaluation criteria differed between preferences for AI-generated and manually created explanations. As Table~\ref{tab:chi_square_explanations} shows, none of the differences were statistically significant ($p > 0.05$). The most frequent reason in both groups was Relevance, followed by Correctness and Clarity. In summary, while participants demonstrated a slight and statistically significant overall preference for manually created explanations, individual choices were highly diverse, and the type of evaluation criterion cited did not differ systematically between manual and AI-generated explanations. The results indicate that the quality of explanations is perceived as highly context-dependent and subjective, with no clear consensus among users.

\begin{table}[ht]
    \centering
    \caption{Chi-Square Test Results for Explanation Preferences}
    \begin{tabularx}{\columnwidth}{l r r r r}
        \toprule
        \textbf{Category} & \textbf{\# ChatGPT} & \textbf{\# Manual} & \textbf{Chi²} & \textbf{p-value} \\
        \midrule
        Relevance & 176 & 199 & 0.03 & 0.8575 \\
        Correctness & 151 & 170 & 0.10 & 0.7534 \\
        Clarity & 140 & 162 & 0.08 & 0.7761 \\
        Level of Detail & 75 & 105 & 0.03 & 0.8602 \\
        Tone & 29 & 43 & 0.01 & 0.9123 \\
        Style & 20 & 29 & 0.00 & 0.9465 \\
        \bottomrule
    \end{tabularx}
    \label{tab:chi_square_explanations}
\end{table}

\subsection{Tool Evaluation}

To assess the efficiency and usability of our tool, we conducted a further study with the same eight participants from the requirement evaluation. The time needed to formulate requirements was measured for both manual creation and tool-supported generation. With the tool, participants in both groups completed the task substantially faster: Group A (manual first, then tool) reduced average time per four reviews from 5:44 to 3:12 minutes; Group B (tool first, then manual) from 2:40 to 5:37 minutes. This equates to a reduction in formulation time of 44.2\% in Group A and 52.5\% in Group B. Regarding confidence in the generated requirements, participants reported similar levels for both approaches: for the tool-supported results, four felt "very confident" and four "confident"; for the manual results, three felt "very confident", three "confident", and two "neutral". Notably, five participants reported equal confidence for both methods, while three felt more confident using the tool. The tool was well received: six participants "fully agreed" and two "rather agreed" that it supported the formulation of explainability requirements. All eight participants "fully agreed" that they would use the tool again, indicating a high level of acceptance. Overall, the tool not only increased efficiency but also achieved high acceptance among requirements engineers, with comparable confidence in the quality of tool-generated and manually created requirements.

\section{Discussion}
\label{sec:discussion}

\subsection{Answering the Research Questions}

\begin{colora}
    \textbf{Answer to RQ1:} Our results show a clear preference among requirements engineers for manually created explainability requirements, mainly due to higher perceived relevance and correctness. While human-authored requirements are favored in most cases, automatically generated requirements are occasionally preferred for specific reviews—especially when explanation needs are explicit and well-structured. Style and tone had little influence on participant choices.
\end{colora}

\begin{colora}
    \textbf{Answer to RQ2:} No strong overall preference was found among participants for either automatically or manually generated explanations. Manually created explanations were often rated higher for correctness and relevance, while automated explanations were sometimes preferred for their stylistic qualities. Overall, neither approach consistently outperformed the other across all criteria.
\end{colora}

\subsection{Interpretation}

Our findings indicate that requirements engineers tend to prefer manually created explainability requirements, primarily due to issues of relevance and correctness. While AI can capture the main aspects of user explanation needs, it often struggles with contextual nuances and may introduce inaccuracies or ambiguities—a limitation likely rooted in the generalization behavior of large language models. The importance of correctness in participant justifications highlights that AI-generated requirements can be error-prone, particularly in less explicit or ambiguous contexts.

However, the fact that AI-generated requirements were sometimes favored—particularly for explicit and well-structured needs—demonstrates their potential as a supportive tool. This suggests that AI models can be useful for drafting requirements, especially when human analysts are available to review and refine the output. As language models continue to evolve, improvements in relevance and correctness (for example through advanced fine-tuning and domain-specific training) may further close the gap between AI- and human-generated requirements.

In contrast, the evaluation of explanations revealed much greater diversity in user preferences. Here, correctness remained the most important factor, but style also played a larger role. AI-generated explanations were sometimes regarded as more structured or readable, making them attractive for user-facing applications. However, the risk of plausible yet incorrect responses remains, especially when explanations are not grounded in system-specific data. This points to the importance of integrating AI-generated content with domain knowledge and validating outputs before presentation to users.

For practitioners, the results highlight the value of a hybrid approach: AI tools can be used to automate and accelerate initial drafts of requirements and explanations, but human oversight remains essential for ensuring accuracy and contextual appropriateness. In routine or repetitive scenarios (such as responding to frequent user inquiries), AI-generated explanations may already be sufficiently robust if subjected to targeted review. In sensitive domains (such as healthcare or law), strict validation is indispensable.

These findings imply that further development of AI tools should focus on improving factual correctness and domain adaptation, rather than purely stylistic enhancements. Mechanisms for human-in-the-loop validation, domain-specific fine-tuning, and post-processing checks may help to mitigate existing weaknesses. Our tool evaluation also confirms that AI assistance can significantly increase efficiency, but should be embedded in workflows that allow for expert intervention when necessary.

Future research should therefore address the integration of system and domain data into AI models, as well as more systematic evaluation in regulated or high-stakes environments. Enhancing the reasoning capabilities of AI for RE, and ensuring traceability and verifiability of generated content, are key areas for further exploration.

\subsection{Threats to Validity}
\label{sec:threats}
We describe the threats to validity faced by this research and the mitigation actions taken to control them within our possibilities. We organize the threats according to the categories described by Wohlin \textit{et al}.~\cite{wohlin2012experimentation}.

\textbf{Construct Validity.}
The construction of our dataset relied on a workshop with four requirements engineers, who developed the gold standard by consensus. Although this process increases consistency, some subjectivity remains unavoidable. All participants were experienced in RE and familiar with the identification of explanation needs in app reviews, which helps mitigate risks of misinterpretation. Nevertheless, the subjective nature of explanation needs and requirements cannot be fully eliminated. Furthermore, our validation is limited to a single application (Spotify), creating a potential mono-operation bias. Applying the approach to additional apps and domains in future studies would help strengthen construct validity.

\textbf{Internal Validity.}
Participant judgments may have been influenced by individual preferences, familiarity with certain phrasing, or cognitive biases. To minimize such effects, we anonymized the origin of artifacts (manual vs. AI-generated) during evaluation. However, as the evaluation relied on subjective feedback and self-reported justifications, internal validity is partly dependent on participant honesty and interpretation. While objective measures can be useful, subjective assessments are particularly relevant for evaluating the perceived utility and quality of explainability requirements and explanations.

\textbf{Conclusion Validity.}
Our results are based on a limited sample size: eight requirements engineers, 14 end users, and 58 annotated app reviews. This limits the statistical power of our conclusions and the ability to detect small effects or rule out random variation. Accordingly, statistical analysis should be interpreted as indicative rather than definitive. The study focused on a detailed qualitative assessment of a single, well-understood use case, enabling in-depth insights but at the cost of broader generalizability. Further studies with larger and more diverse samples are needed to validate and extend our findings.

\textbf{External Validity.}
Our results are based on a specific context: English-language Spotify reviews, evaluated by requirements engineers from a single German company. While Spotify is a widely used application, we cannot assume direct transferability to other domains, languages, or user types (\textit{e.g}., developers, domain experts). The relevance of explainability needs—and the appropriateness of AI-generated artifacts—may differ in other domains or for different types of users (\textit{e.g.}, developers or end-users). The probabilistic nature of ChatGPT also means that repeated runs may yield different outputs, which poses a challenge for reproducibility. Finally, the requirements engineers evaluating the requirements were not Spotify developers, which may limit contextual accuracy, although their expertise in RE supports the overall relevance of the findings.

\subsection{Future Work}

Our study identifies several directions for further research to advance automated explainability in RE:
First, expanding the dataset to include reviews from multiple apps, domains, and languages would allow assessment of generalizability and domain adaptation. Cross-industry studies could help establish the robustness and flexibility of the approach.
Second, enhancing the precision and relevance of AI-generated requirements and explanations remains a key challenge. Future research should investigate advanced prompting techniques (such as few-shot or chain-of-thought prompting) and fine-tuning with domain-specific data to improve model performance, especially for complex or ambiguous feedback.

Third, broader and more diverse participant samples, including requirements engineers from different organizations and regions, as well as additional user groups (such as developers and end-users), would help uncover patterns and preferences related to demographic or professional background. Analyzing correlations between these factors and voting behavior could provide deeper insights.
Fourth, further analysis of which types of explanation needs are best addressed by AI, and under which circumstances human refinement is indispensable, could help tailor hybrid workflows for practical application.
Finally, future studies should explore practical strategies for delivering explainability—such as integrating explanations into user interfaces, automating responses to user inquiries, or incorporating them directly into software requirements. The impact of these approaches on user trust, satisfaction, and system adoption deserves particular attention, especially in regulated or high-stakes environments.
Overall, these avenues can support the development of more reliable, user-centered, and domain-adapted AI tools for explainability in software engineering.

\section{Conclusion} 
\label{sec:conclusion}
The increasing complexity of modern software systems has amplified the need for explainability to foster transparency and user trust. Yet, the task of deriving explainability requirements from user feedback and generating suitable explanations remains challenging. In this work, we investigated the potential of generative language models—specifically ChatGPT—to support and automate this process. To enable a robust evaluation, we established a manually curated reference dataset through a workshop with four experienced requirements engineers. This dataset formed the basis for two user studies: one in which eight additional requirements engineers assessed the quality of explainability requirements, and another in which 14 end users evaluated explanations—each comparing AI-generated artifacts with manually crafted ones. 

We also developed a tool that leverages large language models to automate the summarization of user reviews, the derivation of requirements, and the generation of explanations. Our findings suggest that AI-generated requirements, while helpful as initial drafts, often lag behind manual ones in terms of relevance and correctness. However, for explanations, AI-generated texts occasionally matched or even exceeded human-written ones in style and clarity—though correctness remained a notable limitation. These results highlight the value of a hybrid approach: AI can accelerate the process, but human review remains essential to ensure quality and contextual accuracy. Our tool evaluation supports this view, demonstrating significant efficiency gains without compromising confidence in the results. As the automation of explainability artifacts is still in its early stages, our study offers an initial foundation and points to several avenues for further improvement. Future work should focus on larger, more diverse datasets, domain-specific fine-tuning, and advanced prompt engineering to further enhance the precision and applicability of AI-assisted explainability in software engineering.

\section*{Acknowledgment}
This work was funded by the Deutsche Forschungsgemeinschaft (DFG, German Research Foundation) under Grant No.: 470146331, project softXplain (2022-2025).
We want to thank Phoenix Contact Electronics GmbH (Blomberg, Germany) for conducting the studies at their company.

\bibliographystyle{IEEEtran}
\bibliography{references.bib}

\end{document}